\documentclass[12pt]{article}
\usepackage{amssymb,graphicx,amsmath,array,verbatim}

\numberwithin{equation}{section} 

\def\thline{\noalign{\hrule height 1pt}}
\def\tvline{\vrule width 1pt} 

\newcommand {\beq}{\begin{eqnarray}}
\newcommand {\eeq}{\end{eqnarray}}
\def\p{\partial}
\newcommand{\NF}{N_{\rm F}}
\newcommand{\NC}{N_{\rm C}}
\newcommand{\vs}[1]{\vspace{#1 mm}}
\newcommand{\hs}[1]{\hspace{#1 mm}}
\newcommand{\bpm}{\begin{pmatrix}}
\newcommand{\epm}{\end{pmatrix}}

\newcommand{\tr}{{\rm Tr}}
\newcommand{\vol}{{\rm Vol}}
\newcommand{\D}{\mathcal D}

\newcommand{\be}{\begin{equation}}
\newcommand{\ee}{\end{equation}}
\newcommand{\bea}{\begin{eqnarray}}
\newcommand{\eea}{\end{eqnarray}}
\newcommand{\beann}{\begin{eqnarray*}}
\newcommand{\eeann}{\end{eqnarray*}}
\newcommand{\nn}{\nonumber}
\newcommand{\ba}{\begin{array}}
\newcommand{\ea}{\end{array}}

\begin{document}
\begin{titlepage}

\setcounter{page}{0}
\renewcommand{\thefootnote}{\fnsymbol{footnote}}

\begin{flushright}
{\tt hep-th/0703197} \\
UT-Komaba/07-2\\
TIT/HEP--569\\
TU--787\\
March, 2007
\end{flushright}

\vspace{0mm}
\begin{center}
{\LARGE \bf Statistical Mechanics of Vortices\\
\bigskip
from D-branes and T-duality
}

\vspace{15mm}
{\normalsize\bfseries
Minoru~Eto$^1$, 
Toshiaki~Fujimori$^2$, 
Muneto~Nitta$^3$, \\
 Keisuke~Ohashi$^2$, 
 Kazutoshi~Ohta$^4$
and
 Norisuke~Sakai$^2$}
\footnotetext{ 
e-mail addresses: \tt{
meto(at)hep1.c.u-tokyo.ac.jp,
fujimori(at)th.phys.titech.ac.jp,\\
nitta(at)phys-h.keio.ac.jp, 
keisuke(at)th.phys.titech.ac.jp,
kohta(at)phys.tohoku.ac.jp,
nsakai(at)th.phys.titech.ac.jp}
}

\vskip 1.5em

{ \it $^1$University of Tokyo,  Inst. of Physics
Komaba 3-8-1, Meguro-ku 
Tokyo 153, JAPAN \\[2mm] 
$^2$Department of Physics, Tokyo Institute of 
Technology \\
Tokyo 152-8551, JAPAN  \\[2mm]
$^3$Department of Physics, Keio University, Hiyoshi,
Yokohama, Kanagawa 223-8521, JAPAN
  \\[2mm]
$^4$Department of Physics, 
Graduate School of Science,
Tohoku University\\
Sendai 980-8578, JAPAN
 }

\end{center}

\vspace{7mm}

\centerline{{\bf Abstract}}
\vspace{3mm}
We propose a novel and simple method 
to compute the partition function of 
statistical mechanics of local and semi-local BPS vortices 
in the Abelian-Higgs model and 
its non-Abelian extension 
on a torus.
We use a D-brane realization of the vortices 
and T-duality relation to domain walls.
We there use a special limit where domain walls reduce to gas of
hard (soft) one-dimensional rods for Abelian (non-Abelian) cases. 
In the simpler cases of the Abelian-Higgs model 
on a torus, our results agree with exact results 
which are geometrically derived by an explicit integration 
over the moduli space of vortices.
The equation of state for $U(N)$ gauge theory deviates from van der Waals one, and the second virial coefficient is proportional to $1/\sqrt{N}$, implying that non-Abelian vortices are ``softer" than Abelian vortices. Vortices on a sphere are also briefly discussed.

\end{titlepage}
\newpage

\renewcommand{\thefootnote}{\arabic{footnote}}
\setcounter{footnote}{0}

\section{Introduction}

Knowledge of moduli space structure of solitons is 
important in order to understand 
not only their own dynamics but also non-perturbative 
effects in field theory. 
The  moduli space structure of the solitons signifies 
its topology, metrics and singularities. 
The volume of the moduli space also possesses 
an important meaning and plays an essential role in understanding 
the non-perturbative dynamics. 
Recently 
Nekrasov has shown that the prepotential of ${\cal N}{=}2$ 
supersymmetric gauge theory in $d=3+1$
can be obtained from some statistical partition function,
whose free energy measures the volume of 
the moduli space of Yang-Mills instantons \cite{Nekrasov:2002qd}.
The instanton moduli space is non-compact 
and of course the volume diverges,
but a coefficient of the divergent part gives perturbative 
and non-perturbative
information on supersymmetric gauge theory.
It is also interesting that the partition function 
or the volume of the instantons
is closely related to topological string amplitudes 
on suitable Calabi-Yau manifolds.
Precisely speaking, the instantons, 
which are used for the calculation of the volume,
are not solutions of the equations of motion in supersymmetric gauge theory.
However, in the calculation of the
partition function or the volume, 
the so-called localization theorem says that 
the result does not depend on details of the moduli space structure.
This suggests that we do not need to know 
the exact solutions and metrics 
in order to calculate the volume of the moduli space 
of some class of BPS solitons. 
However, the issue has not been settled except for Yang-Mills 
instantons, since other examples and applications 
have not been investigated yet. 
 For instance, similar method should be applicable to calculate 
the effective (super)potential of 
$d=1+1$ supersymmetric gauge theory 
from the statistical partition function associated with
the volume of the vortex moduli space.

Another direct application of the volume of the moduli 
space is a classical statistical mechanics of the solitons. 
The Gibbs partition function of solitons at finite 
temperature $T$ is given by an integration over a phase space, 
which is a cotangent bundle $T^*{\cal M}$ of the moduli 
space $\cal M$ of solitons: 
\beq
Z=\frac{1}{(2\pi)^D}
\int_{T^*{\cal M}}d^{D}x\, d^{D} p \, e^{-H(\vec{x},\vec{p})/T}.
\eeq
Here $D$ represents the dimensions of the moduli space, 
and $H(\vec{x},\vec{p})$ is a Hamiltonian of the soliton 
system in terms of the moduli parameters. 
If we assume that the solitons are sufficiently diluted 
and interactions can be ignored, we can regard the 
Hamiltonian as quadratic in momenta 
$H(\vec{x},\vec{p})=\frac{1}{2}g^{ij}(x)p_ip_j$ and 
$g^{ij}$ is the (inverse of) metric on ${\cal M}$. 
Then we can explicitly perform the integration over 
the momenta (cotangent direction). 
The partition function becomes 
\beq
Z = \left(\frac{T}{2\pi }\right)^{D/2}
\int_{\cal M}d^Dx \, \sqrt{\det(g_{ij}(x))}
=  \left(\frac{T}{2\pi}\right)^{D/2} \vol({\cal M}),
 \label{eq:Z-vol}
\eeq
which is proportional to the volume of the moduli space 
\cite{Manton:1993tt}.
Many applications and calculations from this point of view 
can be found \cite{Manton:1993tt}--\cite{Romao:2005ph} 
in the case of Abrikosov-Nielsen-Olesen (ANO) vortices 
\cite{Abrikosov:1956sx}, namely vortices in the 
Abelian-Higgs model at the critical coupling 
(the BPS limit).

On the other hand, non-Abelian BPS vortices have been 
recently found \cite{Hanany:2003hp,Auzzi:2003fs} and 
extensively discussed in the (supersymmetric) 
non-Abelian Higgs model, which is the $U(N_{\rm C})$ 
gauge theory coupled with $N_{\rm F} (> N_{\rm C})$ 
fundamental Higgs fields. 
Like (non-commutative) instantons, 
single vortex in 
the $N_{\rm C} = N_{\rm F}$ case  
is found to carry internal moduli 
$\mathbb CP^{N_{\rm C}-1}$ in addition to position moduli.
They can confine monopoles in the Higgs phase 
\cite{Tong:2003pz,Auzzi:2003em} 
and this fact is applied \cite{Shifman:2004dr,Hanany:2004ea}
to show coincidence of the BPS spectra of 
$d=3+1$, ${\cal N}=2$ supersymmetric gauge theory 
and $d=1+1$, ${\cal N}=2$ supersymmetric $\mathbb CP^n$ model.
The moduli space of multiple ($k$-)vortices has been 
conjectured in string theory \cite{Hanany:2003hp} 
and then has been completely determined 
in field theory \cite{Eto:2005yh,Eto:2006pg,Eto:2006cx}, 
which has turned out to be 
some resolution of $k$-symmetric product 
of $\mathbb C \times \mathbb CP^{N_{\rm C}-1}$ \cite{Eto:2005yh}. 
Although its explicit metric is still unknown 
except for a single ($k=1$) vortex,
an integration formula of the K\"ahler potential 
has been found \cite{Eto:2006uw}.
The structure of the moduli space of 
two ($k=2$) vortices has been worked out 
\cite{Hashimoto:2005hi,Eto:2006cx} 
and is applied to non-Abelian duality \cite{Eto:2006dx}, 
and classical dynamics of non-Abelian vortex-strings 
has been clarified \cite{Eto:2006db}.
The extension to a superconformal field theory 
\cite{Tong:2006pa} as well as the Chern-Simons-Higgs 
theory \cite{Aldrovandi:2007nb} has been discussed. 

In this article, we propose a novel and simple derivation 
of the volume of the moduli space of Abelian as well as 
non-Abelian (semi-)local vortices in order to apply to 
the statistical mechanics of vortex gas. 
We utilize a geometric and stringy (D-brane) 
interpretation of the Abelian and 
non-Abelian BPS vortices in supersymmetric gauge theory 
\cite{Hanany:2003hp} and 
use a T-dual relation between the vortices 
and domain walls, 
which was proposed in \cite{Tong:2002hi,Eto:2006mz}.
This interpretation shows us 
a schematic structure of the vortex moduli space
without any detailed information on 
the exact solutions and metrics. 
(See Fig.\,17 of Ref.\,\cite{Eto:2006mz} for the moduli space of
two non-Abelian local vortices in $U(2)$ gauge theory.)
It is interesting to observe 
that the T-duality reduces the calculation of 
the volume into a simple problem of
a gas of hard rods between 
a one-dimensional interval in the Abelian case. 
The volume of the moduli space of the ANO vortices 
in the Abelian-Higgs model (with $N_{\rm C}=N_{\rm F}=1$) 
has been already calculated by 
\cite{Manton:1993tt}--\cite{Manton:1998kq} 
(see also \S{7} in \cite{Manton:2004tk}) 
and an analogy with the hard rod problem was pointed out 
there. 
However this relation has been mysterious for a long time.  
We can explain 
from a point of view of the T-duality in 
superstring theory why the one-dimensional hard rod problem 
appears.  
The advantage of our method is that it can be easily extended 
to the general cases, namely local and semi-local, Abelian 
and non-Abelian vortices. 
We find that the T-duality enables us to reduce the problem 
of a gas of non-Abelian local vortices to a gas of soft 
rods, in contrast to the hard rods in the case of 
Abelian local vortices. 
We obtain that the second virial coefficient is proportional 
to $1/\sqrt{N}$ for non-Abelian local vortices in $U(N)$ 
gauge theories at large $N$. 
This shows that the exclusion volume of a vortex behaves as 
$1/\sqrt{N}$ as we increases $N$, getting closer to an 
ideal gas. Moreover, equation of state for local vortices in $U(N)$ gauge theory deviates from van der Waals one, contrary to the case of the Abelian-Higgs model. 

Comparing with the calculation in 
\cite{Manton:1993tt}--\cite{Manton:2004tk}, 
our result in the Abelian case gives the precise answer 
even though we 
take a special limit 
to the configuration 
and 
do not use the exact metrics. 
So we expect that this is an another example where the localization
theorem effectively works. 
In fact calculation of the volume of 
vortex moduli space using the localization theorem
can be found in the case of 
the Abelian-Higgs model \cite{Romao:2005ph}. 

In Sec.~2 we give a brief review of 
brane configurations of vortices and 
domain walls.
In sec.~3, we calculate the partition function of 
vortices in the Abelian-Higgs model, 
which agrees with the previous result.
In sec.~4, we give general formula 
of the partition function of 
local/semi-local non-Abelian vortices 
which is new and shows the power of our method. 
We perform explicit calculation 
in several cases. 
Sec.~5 is devoted to a discussion. 
An interesting duality is observed and 
our method is extended to the case of a base 
manifold 
$S^2$. 
Appendix gives some details of derivation of the virial 
coefficient.

\section{Non-Abelian Vortices 
and D-brane Interpretation}

We first start with (2+1)-dimensional $U(\NC)$ gauge theory 
with $\NF (\geq \NC)$ massless Higgs fields in
the fundamental representation. 
The Lagrangian is given by
\beq
\mathcal L 
= \tr \left[ - \frac{1}{2g^2} F_{MN} F^{MN} 
+ \D_M H \left( \D^M H \right)^\dagger 
- \frac{g^2}{2} \left( HH^\dagger - c \mathbf 1_{\NC} \right)^2 \right],
\eeq
where $M,N$ are the indices of space-time $(M,N = 0,1,2)$ 
and the space-time metric is chosen as 
$\eta_{MN} = {\rm diag}(+1,-1,-1)$. 
The field strength is defined by 
$F_{MN} \equiv -i [\D_M, \D_N] = \p_M W_N - \p_N W_M + i [W_M,W_N]$, 
where $W_M$ is the $U(\NC)$ gauge field. 
The scaler fields $H^{rA}\,(r=1,\cdots,\NC),\,(A=1,\cdots,\NF)$ 
are expressed by elements of 
an $\NC \times \NF$ matrix and the covariant derivative is 
defined by $\D_M H \equiv \p_M H + i W_M H$.
The constants $g$ and $c\,(>0)$ are the gauge coupling 
constant and the Fayet-Iliopoulos (FI) 
parameter
 respectively.

For static configurations, the energy is bounded from 
below as follows:
\beq
E &=& 
\int d^2 x \, \tr \left[ \frac{1}{g^2} 
\left( F_{12} + \frac{g^2}{2}
( c \mathbf 1_{\NC} - HH^\dagger ) \right)^2 
+ \D_{\bar z} H \left( \D_{\bar z} H \right)^\dagger 
- c F_{12} \right] \nonumber \\
&\geq& - c \int d^2 x \, F_{12} = 2\pi c k,
\label{eq:bound}
\eeq
where the integer 
$k \equiv - \displaystyle \frac{1}{2\pi} \int d^2 x \, F_{12} $ 
represents the number of vortices (vorticity). 
Since the configurations of the BPS vortices saturate the 
inequality Eq.~(\ref{eq:bound}), 
they satisfy the following BPS equations (the vortex equations):
\beq
 \D_{\bar z} H = 0,  
\hspace{1cm}
 F_{12} + \displaystyle \frac{g^2}{2}( c \mathbf 1_{\NC} - HH^\dagger ) = 0. 
 \label{eq:BPS}
\eeq
If the 
fluctuation energy 
above the energy of the 
multi-vortex static configuration  
is less than that of the mass gap between massless moduli 
fluctuations and massive fluctuations, the dynamics of 
the system is well described by the geodesic motion on 
the moduli space $\mathcal M_k$ 
of the $k$ vortices (Manton's moduli/geodesic approximation 
\cite{Manton:1981mp}), 
so that the Lagrangian for multi-vortex system is given by
\beq
L = g_{i \bar j}(\phi,\bar \phi) \dot{\phi}^i \dot{\bar \phi}^j, 
\label{eq:voreff}
\eeq
where 
$\phi^i,\bar \phi^i,\left(~i=1,\cdots,D
={\rm dim}_{\mathbb C} \, \mathcal M_k\right)$ 
are complex coordinates of the moduli space 
$\mathcal M_k$ and $g_{i \bar j}(\phi,\bar \phi)$ is 
a K\"ahler metric of $\mathcal M_k$.

In the cases of compact base manifolds, 
there 
is the maximum number for vortices 
allowed  
to exist \cite{Bradlow:1990ir}.
For general $\NC$ with a torus $T^2$, this maximum number 
can be obtained by taking the trace of 
the second equation of the BPS equation (\ref{eq:BPS}) 
and by integrating over $T^2$ as 
\beq
- 2 \pi k + \NC \frac{g^2 c}{2} A 
= \frac{g^2}{2} \int d^2x \, \tr
\left(HH^\dagger \right) \geq 0 \quad 
\leftrightarrow \quad 
A\ge k B_{\NC},
\label{eq:Bradlow}
\eeq 
where $A$ is the area of the torus and we have defined the Bradlow area
\beq
B_{\NC} \equiv \frac{1}{\NC} {4 \pi \over g^2 c}.
\eeq
The limit to saturate this inequality is called the Bradlow limit.
In particular, we can regard $B_1= {4 \pi \over g^2 c}$
as an effective area of the Abelian (ANO) vortex. 
Roughly speaking, the inequality (\ref{eq:Bradlow}) implies 
that we can 
squeeze as many vortices as $N_{\rm C}$ times the maximal 
number of the Abelian ($N_{\rm C}=1$) vortices for a given 
area $A$ of the torus. 
This behavior can be understood intuitively as due to the 
freedom for vortices to avoid occupying the same points 
in their additional internal moduli space 
$\mathbb{C}P^{\NC-1}$ when they are squeezed too much in 
the actual configuration space. 
This freedom allows the non-Abelian vortices to overlap 
in the configuration space more easily compared to Abelian 
vortices. 
Therefore the effective area of the non-Abelian vortex $B_{\NC}$
becomes $1/N_{\rm C} < 1$ times the area
of the Abelian vortices $B_1$. 
The Bradlow area $B_{N_{\rm C}}$ indicates 
the smallness of the effective area of vortices at the 
limiting high density of vortices. 
Note that  
there exists another inequality 
in the case of non-Abelian 
gauge theories 
with $N_{\rm C}> k$. 
Since all vortex solutions in this case can be embedded to 
a theory with a smaller number of colors 
$N_{\rm C}'(=k')=k$, the 
vortex configuration for $N_{\rm C}>k$ should satisfy the 
inequality for $N_{\rm C}'=k$,  
that is,     
\beq
 A  \geq {4 \pi \over g^2 c} ~~ (= B_1)
  \label{eq:another_ineq}
\eeq
stating that the area of the torus must be 
larger than that of one ANO vortex.

As we mentioned in Eq.~(\ref{eq:Z-vol}), the classical 
partition function of $k$-vortex system is proportional 
to the volume of 
the  
moduli space $\mathcal M_k$. 
\begin{figure}[h]
\begin{center}
\begin{tabular}{ccc}
\includegraphics[width=50mm]{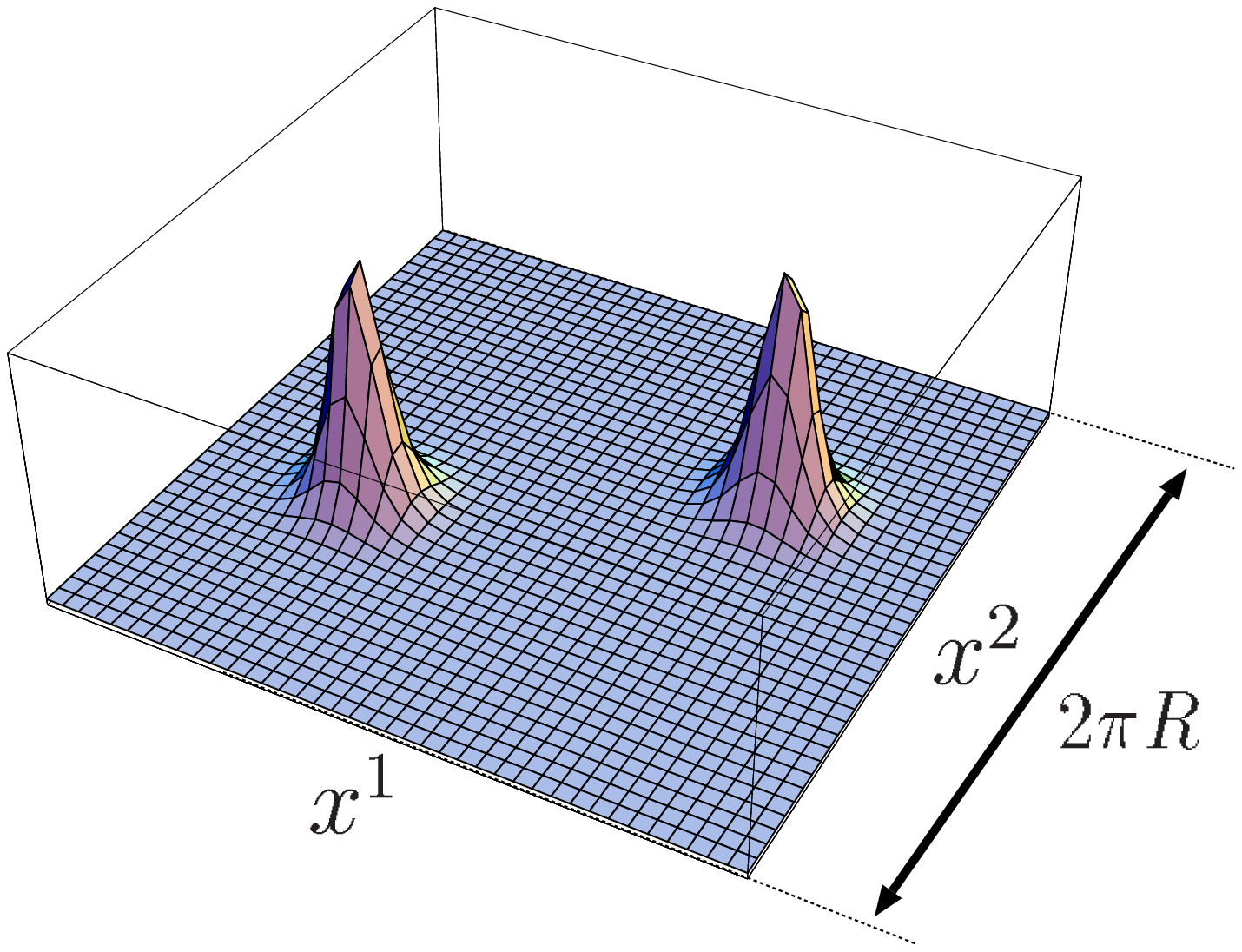} &
\includegraphics[width=50mm]{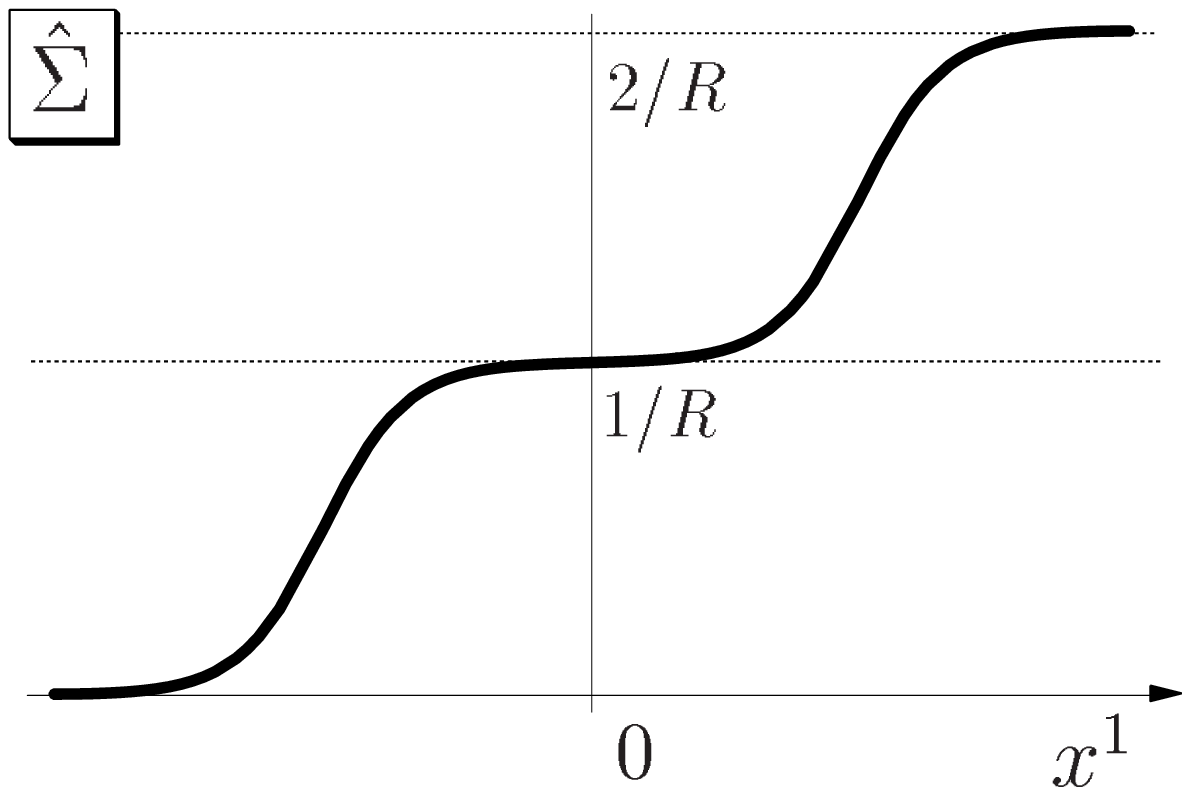} &
\includegraphics[width=50mm]{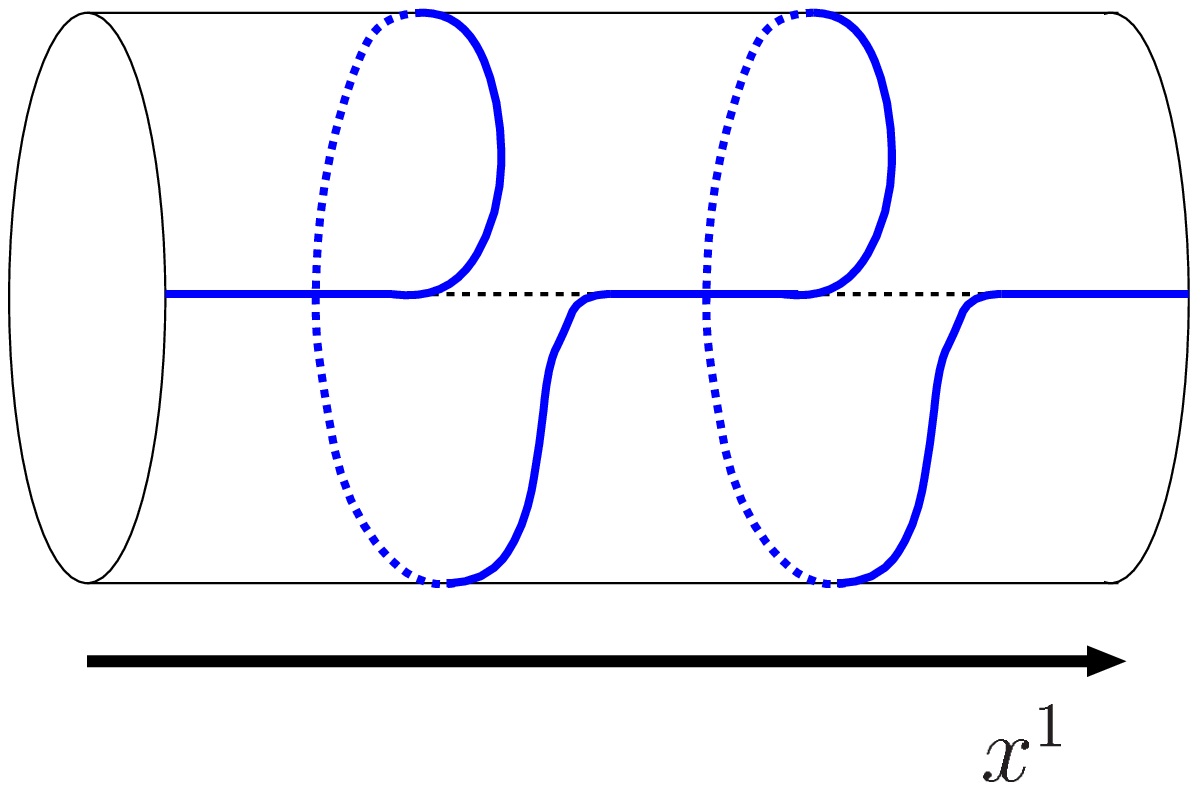} \\
(a) & (b) & (c)
\end{tabular}
\caption{\small Duality between vortices and domain walls: 
$\NC=\NF=1,~k=2$. 
(a) Profile of energy density of 2-vortex configuration on 
$\mathbb R \times S^1$ with period $2 \pi R$, 
(b) Profile of 
$\hat \Sigma(x^1) = -\frac{1}{2\pi R} \int dx^2 W_2$, 
(c) Profile of $\hat \Sigma(x^1)$ written on 
$\mathbb R \times S^1$ with period $1/R$.}
\label{fig:dual}
\end{center}
\end{figure}
To calculate this partition function (volume), we utilize 
the duality relation between vortices and domain walls 
\cite{Tong:2002hi,Eto:2006mz}. 
If we compactify $x^2$-direction on $S^1$ as 
$x^2 \sim x^2 + 2 \pi n R$, a vortex configuration can be 
viewed as domain walls. 
The profile of the kink solution of the domain walls is 
described by the eigenvalues of $\hat \Sigma$ defined by
\beq
\hat \Sigma(x^1) 
= - \frac{1}{2\pi i R} 
\log \left[ \mathbf P \exp 
\left( i \int_0^{2\pi R} dx^2 \, W_2(x^1,x^2) \right) \right],
\eeq
where $\mathbf P$ stands for the path-ordered product.
Note that under the gauge transformation 
$U = e^{-i n x^2/R} \mathbf 1_{\NC}$ ($n \in \mathbb Z$), 
$\hat \Sigma(x^1)$ transforms as 
$\hat \Sigma \rightarrow \hat \Sigma + n/R \mathbf 1_{\NC}$ 
and there is an identification 
$\hat \Sigma \sim \hat \Sigma + n/R \mathbf 1_{\NC}$. 
Thus the eigenvalues of $\hat \Sigma(x^1)$ are interpreted 
as a function which takes value in $S^1$ with radius $1/R$. 
Fig.~\ref{fig:dual} depicts an example of the $\NC=\NF=1$ 
case. 
The $k$-vortex configuration (Fig.~\ref{fig:dual}-(a)) 
corresponds to the $k$-wall configuration 
(Fig.~\ref{fig:dual}-(b)) in the fundamental region, 
and it also can be interpreted as the configuration of 
$\hat \Sigma(x^1)$ with $k$ windings 
(Fig.~\ref{fig:dual}-(c)) on the cylinder.

This duality can be regarded as a T-duality between 
the brane configurations of vortices \cite{Hanany:2003hp} 
and domain walls \cite{Eto:2004vy}.
Our model can be embedded into the supersymmetric system with eight  
supercharges and the associated
model is realized by a combination of various kinds of branes in Type  
IIB superstring theory.
The brane configurations are expressed in Table 1 and drawn in Fig.2  
schematically. 
\begin{table}[h]
\begin{center}
\begin{tabular}{!{\tvline}c!{\tvline}cccccccccc!{\tvline}}
\thline
&$x^0$&$x^1$&$x^2$&$x^3$&$x^4$&$x^5$&$x^6$&$x^7$&$x^8$&$x^9$\\
\thline
$\NC$ $\rm D3$&$\bullet$&$\bullet$&$\bullet$&$\bullet$
&$-$&$-$&$-$&$-$&$-$&$-$ \\
\hline
$\NF$ $\rm D5$&$\bullet$&$\bullet$&$\bullet$&$-$&$\bullet$
&$\bullet$&$\bullet$&$-$&$-$&$-$ \\
\hline 
$2\hs{2}{\rm NS}5$&$\bullet$&$\bullet$&$\bullet$&$-$&$-$&
$-$&$-$&$\bullet$&$\bullet$&$\bullet$ \\
\hline
$k\hs{2}\rm D1$&$\bullet$&$\times$&$\times$&$-$&$\bullet$
&$-$&$-$&$-$&$-$&$-$ \\\thline 
\end{tabular} \\
\vs{5} (a) Brane configuration for vortices \\ \vs{10}
\begin{tabular}{!{\tvline}c!{\tvline}cccccccccc!{\tvline}}
\thline
&$x^0$&$x^1$&$x^2$&$x^3$&$x^4$&$x^5$&$x^6$&$x^7$&$x^8$&$x^9$\\
\thline
$\NC$ $\rm D2$&$\bullet$&$\bullet$&$-$&$\bullet$&
$-$&$-$&$-$&$-$&$-$&$-$ \\
\hline
$\NF$ $\rm D4$&$\bullet$&$\bullet$&$-$&$-$&$\bullet$&
$\bullet$&$\bullet$&$-$&$-$&$-$ \\
\hline 
$2\hs{2}{\rm NS}5$&$\bullet$&$\bullet$&$\bullet$&
$-$&$-$&$-$&$-$&$\bullet$&$\bullet$&$\bullet$ \\
\hline
$k\hs{2}\rm D2'$&$\bullet$&$\times$&$\bullet$&$-$&$\bullet$&
$-$&$-$&$-$&$-$&$-$ \\
\thline
\end{tabular} \\
\vs{5} (b) Brane configuration for domain walls 
\end{center}
\caption{\small Brane configurations for vortices and 
domain walls: Branes are extended along directions denoted 
by $\bullet$, and are not extended along directions 
denoted by $-$. 
The symbol $\times$ denotes the codimensions of the k 
D1-branes (D2'-brane) on the worldvolume of the D3-branes 
(D2-branes) excluding the $x^3$-direction which is a finite 
line segment.}
\label{tab:brane}
\end{table}
\begin{figure}[h]
\begin{center}
\begin{tabular}{cc}
\includegraphics[width=80mm]{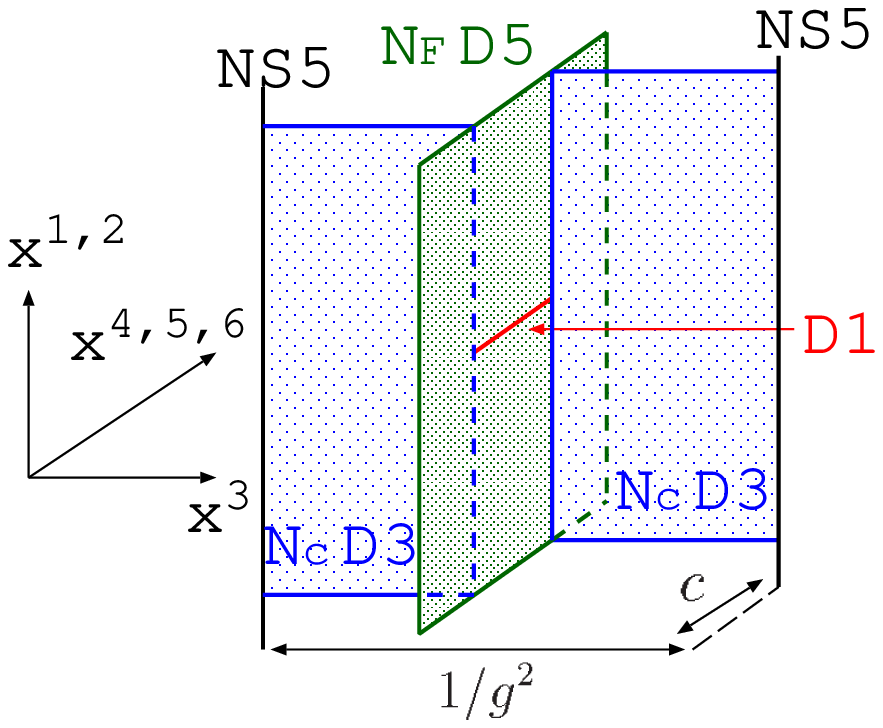} &
\includegraphics[width=80mm]{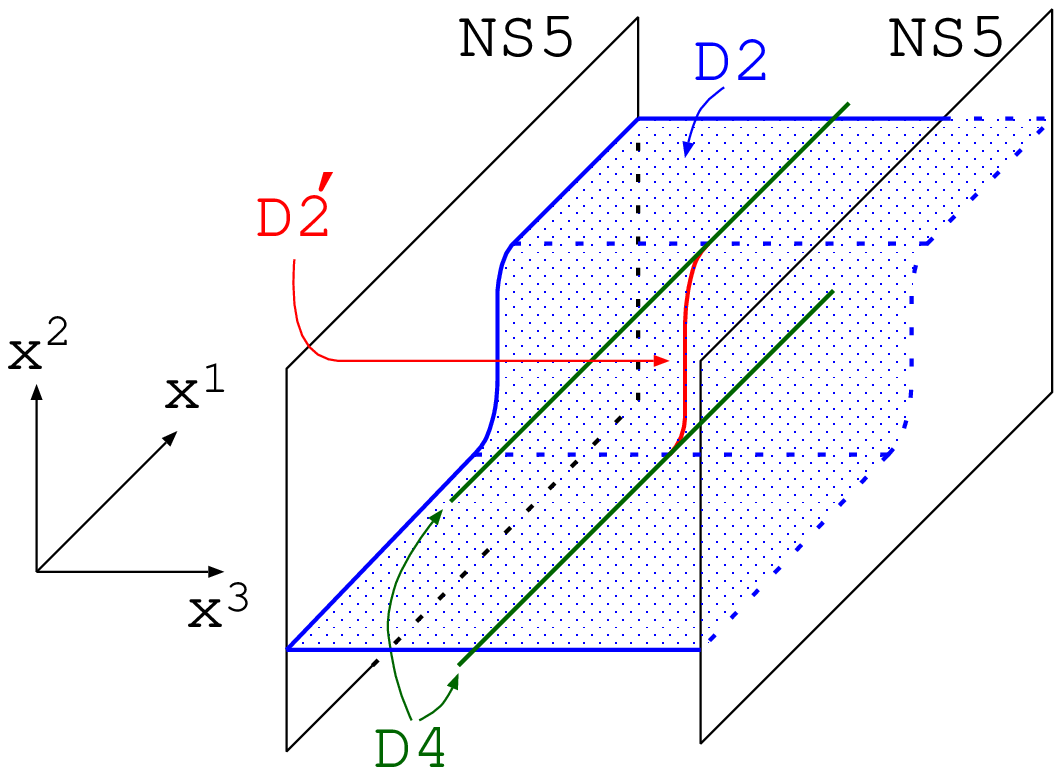} \\
(a) Brane configuration for vortices 
& (b) Brane configuration for domain wall 
\end{tabular}
\end{center}
\caption{\small Brane configurations for vortices and domain wall}
\label{fig:test}
\end{figure}
The $(2+1)$-dimensional $U(\NC)$ gauge theory 
coupled with $\NF$ massless hypermultiplets is 
realized on $\NC$ $\rm D3$ branes in 
the Hanany-Witten setup \cite{Hanany:1996ie} 
as in Fig.~\ref{fig:test}-(a),
since the $x^3$ direction of the $\rm D3$ brane worldvolume 
is a finite line segment.   
$\rm D1$ branes correspond to vortices 
in the $\rm D3$ brane worldvolume theory, 
since they are interpreted 
as codimension-two objects on the $\rm D3$ branes 
\cite{Hanany:2003hp}. 
Fig.~\ref{fig:test} shows the brane configuration 
T-dualized along the $x^2$-direction. 
The theory without vortices is mapped 
to the world-volume theory on the $\NC$ D2 branes, 
which is the (1+1) dimensional $U(\NC)$ gauge theory 
with $\NF$ (massive) hypermultiplets. 
The vortices are mapped to the kinky D2-branes 
representing domain walls \cite{Lambert:1999ix,Eto:2004vy}. 
The eigenvalues of $\hat \Sigma(x^1)$ can be 
interpreted as the position of D2 branes in $S^1$.

\section{A Limit of the Profile and the Moduli Integration: 
the Abeain-Higgs Model}

If we use the T-dual relation via the brane configuration, 
the evaluation of the volume of the vortex moduli space 
reduces to a calculation of the volume of the domain wall 
configurations (kink profiles).
However, it is difficult to solve the domain wall equations 
and to integrate over their configuration space in general.

In order to proceed the evaluation, we demand an 
approximation which 
simplifies  
the profile of the domain wall solution.
Now, let us consider a special limit 
\begin{eqnarray} 
 g^2c, \; 1/R \rightarrow \infty, \quad \text{with} \quad
 d \equiv \frac{2}{g^2cR}:  \mbox{ fixed} .
  \label{eq:lim}
\end{eqnarray} 
Of course, this is a rough approximation with respect to 
the kink profile, we will amazingly find that we can 
obtain the exact results even in this limit. 
Note that using $d$ defined above the inequalities 
(\ref{eq:Bradlow}) and (\ref{eq:another_ineq}) 
are translated in the T-dualized picture to 
\beq
2 \pi R' \geq \frac{1}{\NC} k d\ , \qquad  
    2 \pi R' \geq d, \label{eq:Bradlow2}
\eeq
respectively. 
These inequalities can be easily shown even in this 
T-dualized picture as dicussed below. 
This fact is the first evidence for the usefulness of our 
T-dualized picture 
in this paper.

\begin{figure}[h]
\begin{center}
\begin{tabular}{cc}
\includegraphics[width=55mm]{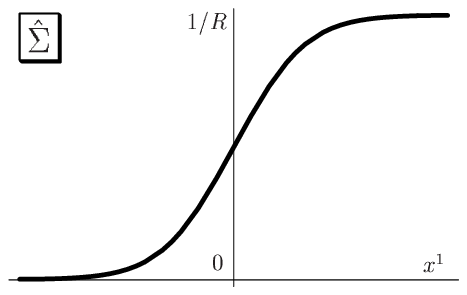} &
\includegraphics[width=55mm]{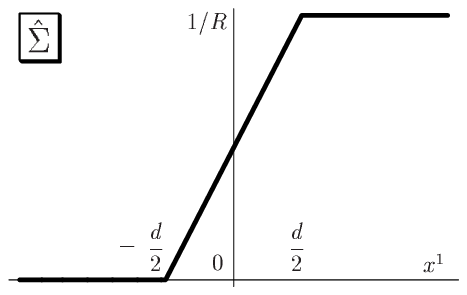} \\
(a) & (b)
\end{tabular}
\caption{\small Profiles of $\hat \Sigma(x^1)$ before and after taking the limit $g^2c,1/R \rightarrow \infty$ with $d \equiv 2/g^2cR$ finite.}
\label{fig:limit}
\end{center}
\end{figure}

Here we describe a limit shape of 
the ANO vortices in the Abelian-Higgs model 
($\NC=\NF=1$) for simplicity.
In this case, the moduli space of 
the multi-vortex system consists of the position of each vortex.
The positions of vortices in the $x^2$-direction are 
translated into the phase degrees of freedom of domain walls
after the T-duality along $x^2$\-direction. 
If one vortex is located at $z=0$,
the corresponding function $\hat \Sigma(x^1)$ is given by
\beq
\hat \Sigma(x^1) = \left\{ \begin{array}{clc} 
0&\quad\text{for}\quad & x^1 < - \dfrac{d}{2} \\ 
\dfrac{1}{Rd} \left( x^1 + \dfrac{d}{2} \right) & \quad\text{for}\quad & 
- \dfrac{d}{2} < x^1 < \dfrac{d}{2} \\
\dfrac{1}{R} &\quad \text{for} \quad& x^1 > \dfrac{d}{2} \end{array} \right..
\eeq
From this expression, we find that the parameter $d$ represents the effective thickness of the dual domain wall. Fig.~\ref{fig:limit} shows the profiles of $\hat \Sigma(x^1)$ before (Fig.~\ref{fig:limit}-(a)) and after (Fig.~\ref{fig:limit}-(b)) taking the limit.
We have previously shown both from field theory \cite{Isozumi:2004jc}
and string theory (the $s$-rule) \cite{Eto:2004vy}
that these domain walls cannot pass through nor 
be compressed with each other. 
Therefore these objects can be regarded as 1-codimension 
rigid bodies with length $d$.
Using this fact, 
we can map the multi-vortex system to the gas of hard rods, 
namely 1-dimensional gas of the rigid bodies.

\begin{figure}[h]
\begin{center}
\begin{tabular}{cc}
\includegraphics[width=60mm]{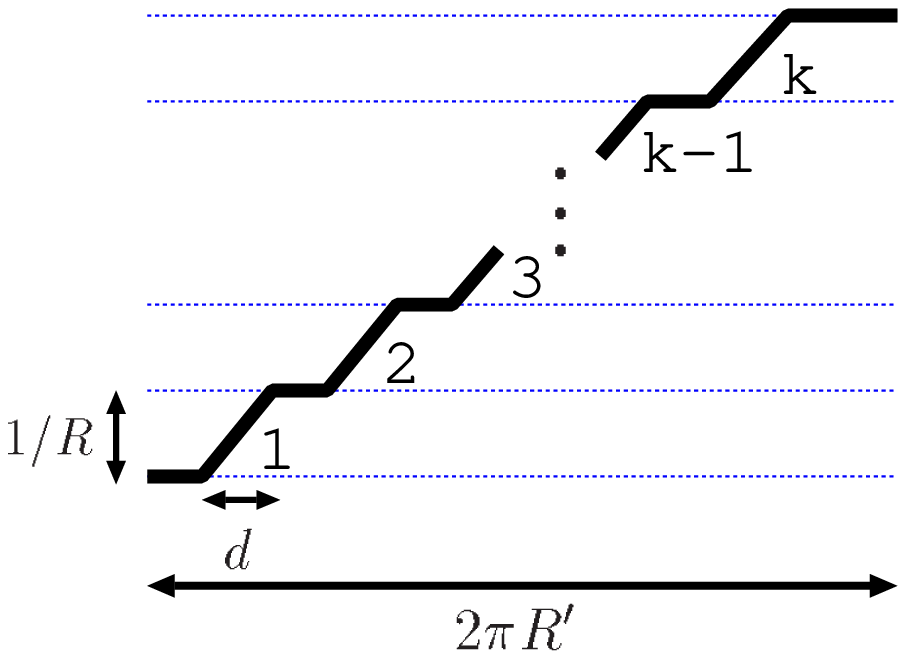} &
\includegraphics[width=60mm]{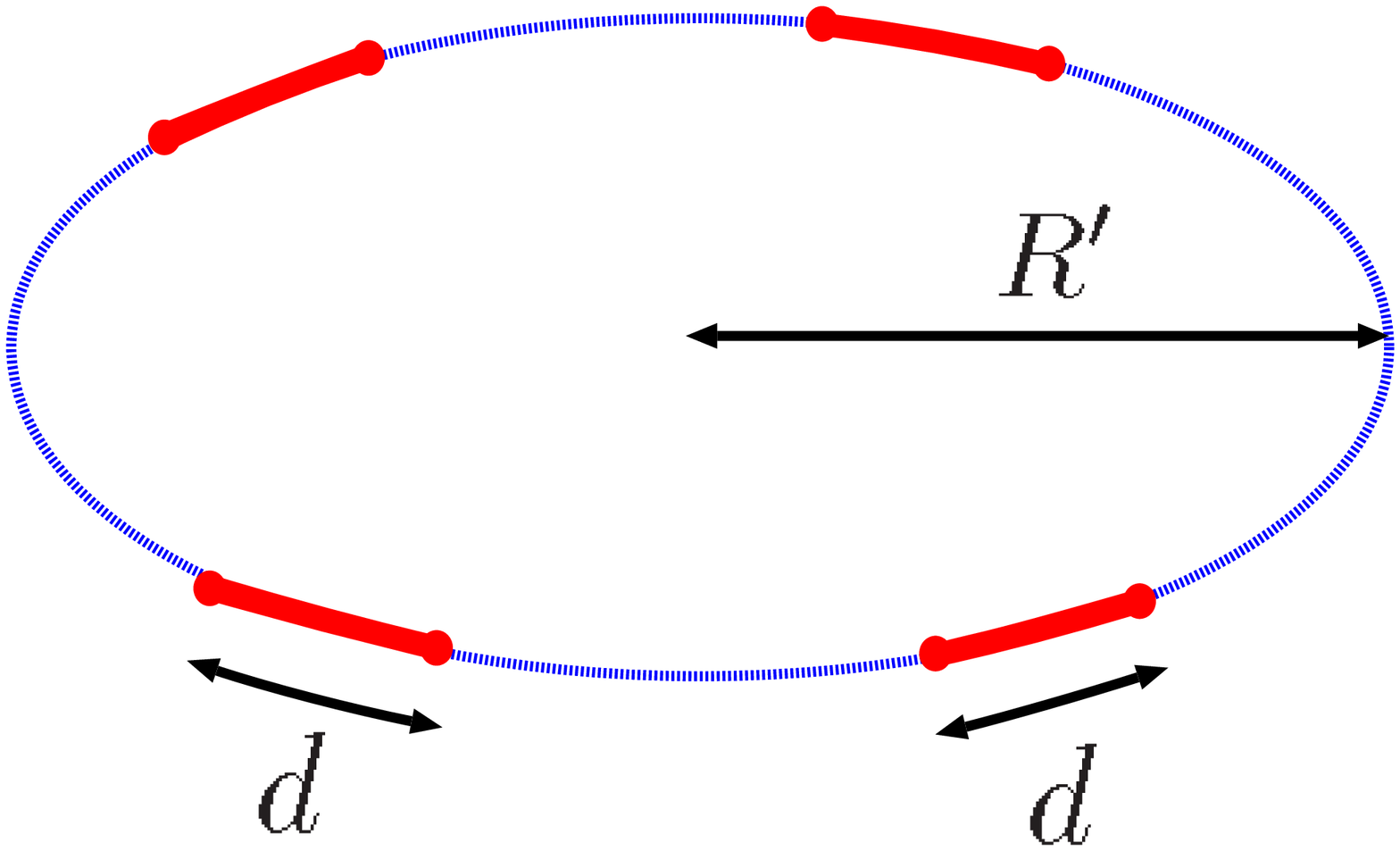} \\
(a) Profile of $\hat \Sigma(x^1)$ & (b) 1-d gas of hard rods on $S^1$
\end{tabular}
\end{center}
\caption{\small Profile of $\hat \Sigma(x^1)$ for vortices and 
corresponding 1-d gas of hard rods.}
\label{fig:vorticesontorus}
\end{figure}

Using the above limit, where the domain walls are regarded as the hard rods in 1-dimension, let us now consider the vortices on a torus $T^2$ 
with periods $2\pi R$ and $2\pi R'$ and dual domain wall configuration space. 
Fig.~\ref{fig:vorticesontorus}-(a) shows the profile of $\hat \Sigma(x^1)$ 
and Fig.~\ref{fig:vorticesontorus}-(b) shows 
the corresponding system of hard rods. 
For the gas of identical hard rods with mass $m$ on $S^1$ with period $L$, 
the classical partition function can be easily calculated as
\beq
Z_{\rm rods} = \frac{1}{k!} \left( \frac{m T}{2\pi} \right)^k L \left( L - kd \right)^{k-1}.
\eeq
In this case, the period is $L = 2 \pi R'$ and each rod has mass $m = 2 \pi c$ which corresponds to the mass of the vortex. There are additional phase degrees of freedom which correspond to the positions of vortices in $x^2$-direction. In the limit $g^2c, 1/R \rightarrow \infty$ with $d=\frac{2}{g^2 c R}$ finite, these phase degrees of freedom become independent and have the period $2\pi R$. 
In other words, the T-duality maps the FI-parameter $c$ to $2\pi R c$. So we should replace $c$ by $2\pi R c$ in the
vortex picture.
(Note that the combination $d=\frac{2}{g^2 c R}$ is invariant under the T-duality.)
Therefore the partition function of $k$-vortex system 
on a torus $T^2$ is given by
\beq
Z^{\NC=\NF=1}_{k,T^2} 
= \frac{1}{k!} \left(c T \right)^k 2 \pi R' \left( 2 \pi R' - kd \right)^{k-1} \left( 2\pi R \right)^k = \frac{1}{k!} \left(c T \right)^k A \left( A - \frac{4 \pi k}{g^2c} \right)^{k-1},
\label{eq:Z_k_ANO}
\eeq
where $A = (2\pi)^2 R R'$ is the area of $T^2$. 
This result coincides with the exact partition function 
which has already been known 
\cite{Shah:1993us,Manton:1998kq,Manton:2004tk}.
The inequality (\ref{eq:Bradlow}) with $\NC=1$ implies that 
the number of vortices $k$ must be less than 
$\frac{g^2 c}{4\pi} A$. 
In the Bradlow limit $k = \frac{g^2 c}{4\pi} A$, 
the partition function Eq.~(\ref{eq:Z_k_ANO}) vanishes. 
In our interpretation of vortices as 1-dimensional rods, 
this maximum number can be understood from the fact that 
sum of the length of rods can not exceed the compact period 
of $S^1$, namely $ \frac{2k}{g^2c R} = k d \leq 2 \pi R'$ 
from the first equation of (\ref{eq:Bradlow2}).

We can derive the van der Waals equation of state for the 
vortex gas in the thermodynamic limit 
$A\to\infty$, $k\to \infty$ with keeping $k/A$ fixed (see eg. \S7.15 in \cite{Manton:2004tk}),
\beq
P\left(A-\frac{4\pi k}{g^2 c}\right)= k T .
\label{eq:eos}
\eeq
Here $P$ is the pressure of the vortex gas and the Boltzman constant is unity $(k_B=1)$ in our notation.
From this we find the pressure of the vortex gas 
diverges at the maximal number of vortices determined by
the area and couplings.
Compared with the general van der Waals equation of state
$\left(P+a k^2/A^2\right)\left(A-kb\right) = kT$ we have 
$a=0$ and $b= 4\pi/(g^2c)$.
The former can be understood as the BPS property 
that there exists no potential energy between BPS vortices. 
The latter implies the size
(exclusion area)
of the ANO vortex is $4\pi/(g^2c)$.

\section{Local/Semi-local Non-Abelian Vortices}

\begin{figure}
\begin{center}
\includegraphics[width=130mm]{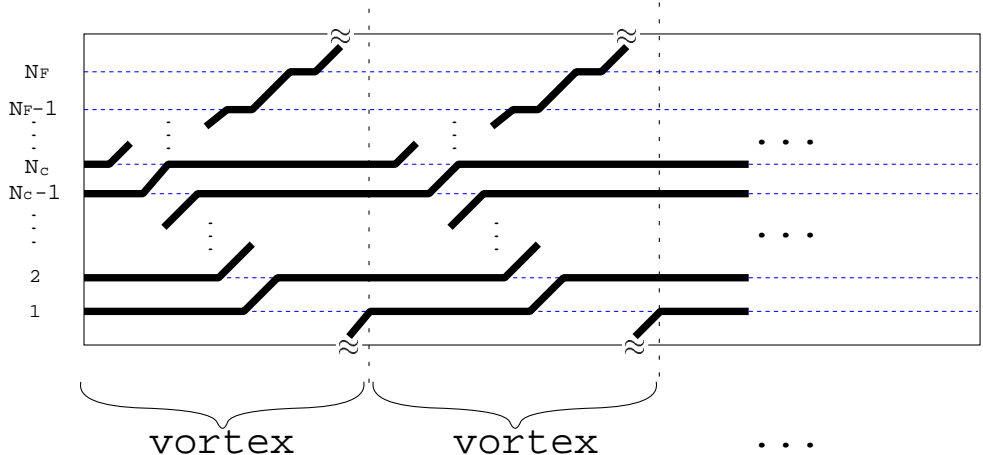}
\caption{\small 
Eigenvalues of $\hat \Sigma(x^1)$ 
as the dualized configuration of $k$ vortices. 
Kinks represent $kN_{\rm F}$ domain walls. 
}
\label{fig:k-vortex}
\end{center}
\end{figure}

\begin{figure}[h]
\begin{center}
\begin{tabular}{cc}
\includegraphics[width=70mm]{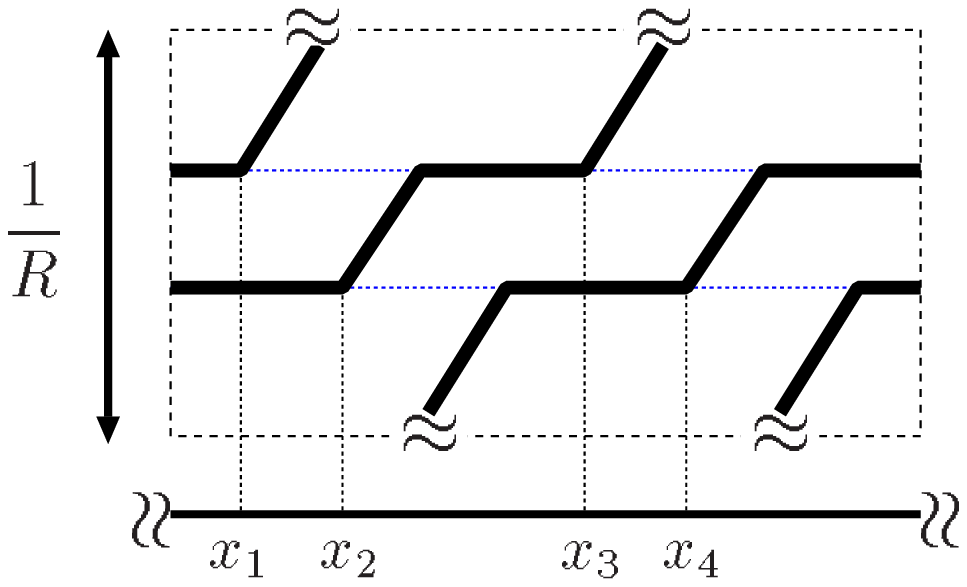} &
\includegraphics[width=74mm]{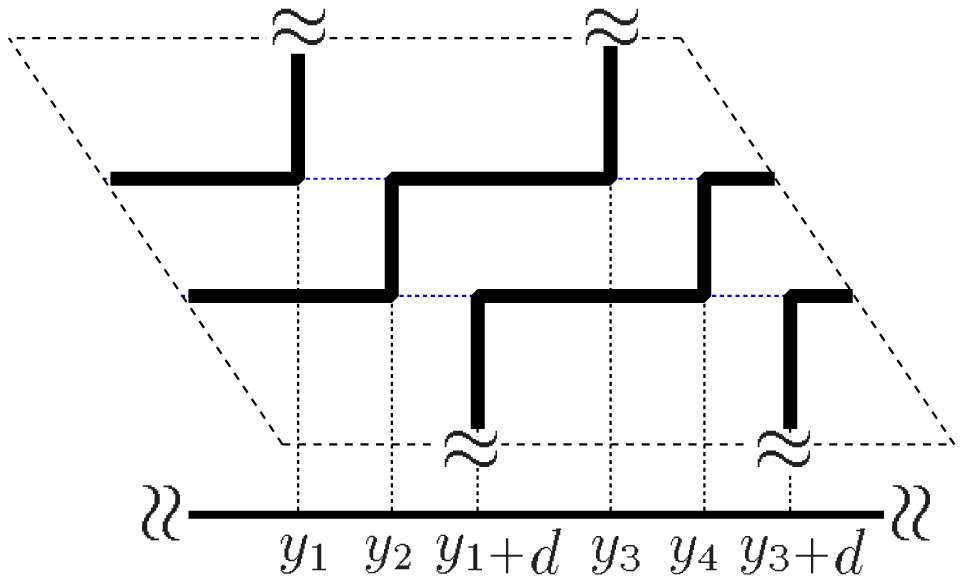} \\
(a) Configuration for $\NC=\NF=2,k=2$ & (b) 
Slant torus 
configuration
\end{tabular}
\end{center}
\caption{\small The profile of eigenvalues of $\hat \Sigma(x^1)$ and slant configuration for $\NC=\NF=2, k=2$. }
\label{fig:nc2nf2k2}
\end{figure}

\begin{figure}[h]
\begin{center}
\begin{tabular}{cc}
\includegraphics[width=60mm]{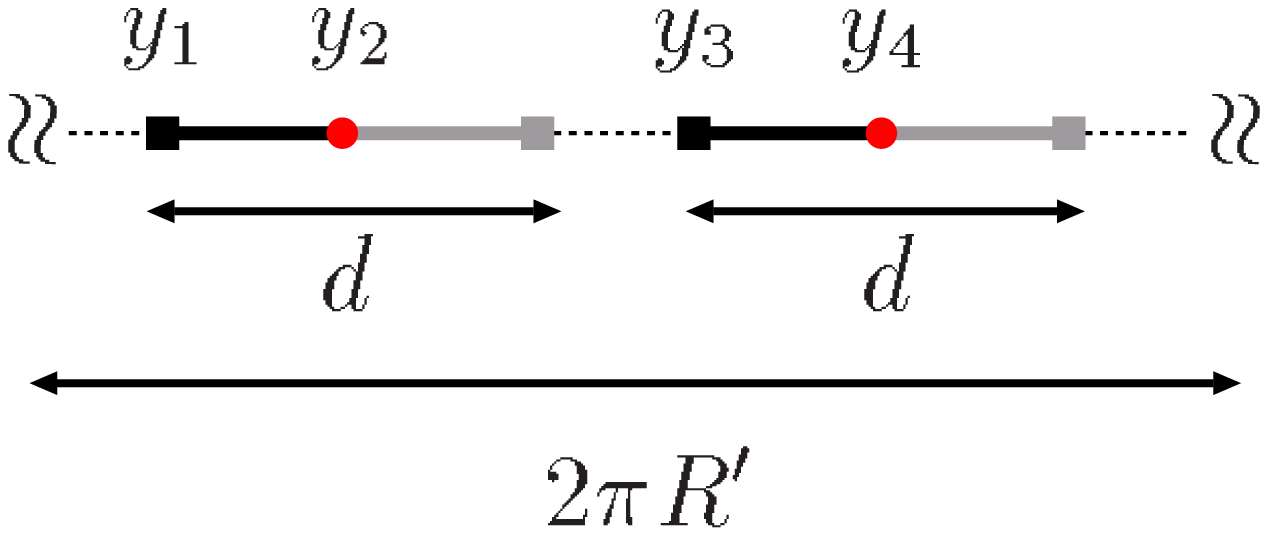} \hs{5} &
\includegraphics[width=60mm]{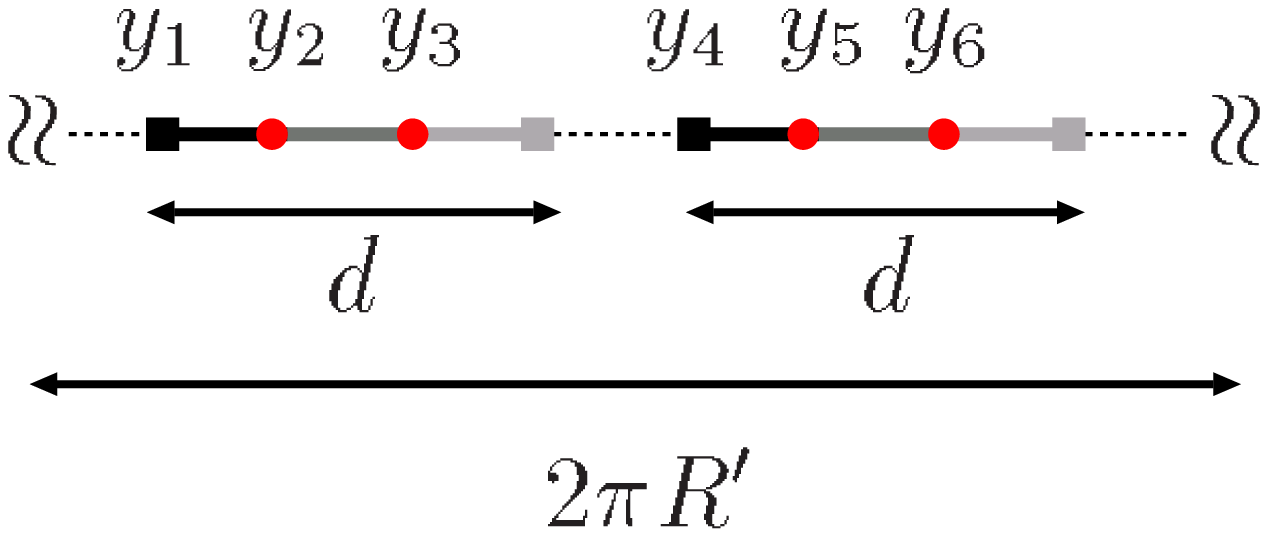} \\
(a) $\NC=\NF=2,k=2$ \hs{5} & (b) $\NC=\NF=3,k=2$ 
\end{tabular}
\end{center}
\caption{\small 
(a) Configuration of rods ($_{\blacksquare}\!\!\!-\!\!\!-\!_{\blacksquare}$)
and particles ($\bullet$) pierced by the rods for $\NC=\NF=2,k=2$
corresponding Fig.~\ref{fig:nc2nf2k2}. 
(b) Those for $\NC=\NF=3,k=2$. 
Each rod is divided by particles into the $N_{\rm C}$ parts 
with different colors.
Each part of rods can overlap with parts of the other rods 
with different colors but not with parts with the same color.
}
\label{fig:nc2nf2k2r}
\end{figure}

So far we have seen that the partition function 
for multi-vortex system can be calculated by using a system 
of hard rods in the case of the Abelian-Higgs model 
($\NC=\NF=1$). 
We next extend this method to the more general cases 
of non-Abelian and Abelian gauge 
theories: 
local non-Abelian vortices 
($N_{\rm C} = N_{\rm F} >1$) and 
semi-local (non-Abelian) vortices ($N_{\rm F} > N_{\rm C}$).
These cases were not known previously, 
which proves the power of our method. 
To this end, we treat vortices in a model 
with a twisted boundary condition 
\beq
H(x^1,x^2+2\pi R) = H(x^1,x^2)e^{2\pi i R M}, 
 \label{eq:twist}
\eeq
where 
$M = {\rm diag} (\alpha_1/R,\cdots,\alpha_{\NF}/R),~ 
\alpha_1<\cdots<\alpha_{\NF}<1+\alpha_1$. 
We can reproduce the partition function with ordinary 
boundary condition by taking the limit 
$\alpha_A \rightarrow 0$ after calculating the partition 
function with the twisted boundary condition. 
In this case, $k$-vortex configuration corresponds to 
$k \NF$ domain walls 
represented as kinks of the eigenvalues 
of $\hat \Sigma(x^1)$ 
as shown in Fig.~\ref{fig:k-vortex}.
By taking the limit (\ref{eq:lim}), we can identify each 
domain wall with a 1-dimensional rod as before. 
In this limit, all 
kinks of $\hat \Sigma(x^1)$ which correspond to 
domain walls have the same slope,  
which is $1/dR = g^2 c/2$. 
In this case there are $\NF$ types of rods whose masses 
are given by 
\beq
m_A &=& 2\pi c (\alpha_{A+1}-\alpha_A), \hs{5} 
A= 1, \cdots, \NF-1, \nonumber \\m_{\NF} 
&=& 2\pi c(1+\alpha_1-\alpha_{\NF}).
\eeq
and the period of phases are given by 
\beq
l_A &=& 2\pi R/(\alpha_{A+1}-\alpha_A),\hs{5} A= 1, 
\cdots, \NF-1, \nonumber \\
l_{\NF} &=& 2\pi R/(1+\alpha_1-\alpha_{\NF}).
\eeq 
Then we can calculate the partition function for the 
multi-vortex system in terms of the 1-dimensional hard 
rods. 
Note that we have to take into account the fact that 
there are $k$ indistiguishable sets of rods corresponding 
to $k$ indistinguishable vortices. 
Therefore the partition function should be divided by $k$ 
after integrating over the configuration space of 
distinguishable rods. 
Then the partition function for multi-vortex system takes 
the form of 
\beq
Z^{\NC,\NF}_{k,T^2} 
&=& \left( \frac{T}{2\pi} \right)^{k\NF} 
\left(\prod_{A=1}^{\NF} m_A l_A \right)^k 
\frac{1}{k} \int_X dx_1 \cdots dx_{k\NF} \nonumber \\ 
&=& \left( \frac{T}{2\pi} \right)^{k\NF} 
(2\pi c)^{k\NF} (2\pi R)^{k\NF} 
\frac{1}{k} \int_{X'} dy_1 \cdots dy_{k\NF}.
\label{eq:partition function}
\eeq
Here the integration with respect to positions $\{x_n\}$ 
is taken over the configuration space of the rods, 
which is denoted as $X$. 
In the last line, 
we have redefined the coordinates as $y_n=x_n-\alpha_A d$ 
if the $n$-th rod sits between the $A$-th flavor and the
($A+1$)-th flavor, and 
the corresponding domain of integration is denoted as $X'$. 
The redefinition of the positions 
corresponds to consider a slant torus 
(Fig.~\ref{fig:nc2nf2k2}) where domain wall configurations are 
right-angled and perpendicular to the $x^1$-direction. 
Therefore we can associate the configurations of vortices 
with that of rods with length $d$ and particles with zero 
size pierced by the rods.
Recalling the original profile of the kink solution, we can see there exist some overlapping rules of the rods;
a part of rods which is divide by the particles can overlap with each other, but not be allowed for some of these. 
So we also introduce colored rods in order to clarify the exclusion rules of rods.
 (See Fig.~\ref{fig:nc2nf2k2r}.)
Note that the above formula in the last line is 
completely independent of the parameters $\{\alpha_A\}$. 
Therefore, our result with the twisted boundary condition 
(\ref{eq:twist}) (non vanishing $\alpha_A$) is applicable 
to the ordinary case with non-twisted boundary condition 
($\alpha_A=0$). 

The partition function Eq.\,(\ref{eq:partition function}) can be interpreted as the asymptotic form of partition function of vortex gas on the rectangular torus in the limit Eq.\,(\ref{eq:lim}). Let us assume that the partition function is independent of details of the torus and depends only on the area of torus $A$. This is the case for the partition function in the model with $\NC=\NF=1$. Then we can calculate the partition function for vortex gas on a surface which is topologically a rectungular torus with area $A$. In addition, we can show that Eq.\,(\ref{eq:partition function}) gives the exact form of partition function as follows: The limit Eq.\,(\ref{eq:lim}) can be rewritten as 
\beq
R = \frac{R_0}{\mu},\hs{5} c = \mu c_0,\hs{5} \mu \rightarrow \infty,\hs{5} R_0,c_0, g: {\rm fixed}.
\eeq
From the explicit expression of the metric of moduli space \cite{Eto:2006pg} and dimensional analysis, we can show that the partition function 
takes the form of
\beq
Z^{\NC,\NF}_{k,T^2} &=& (c T)^{k\NF} A^{k\NF} f(g^2c A) \nonumber \\
&=& (c_0 T)^{k\NF} {A_0}^{k\NF} f(g^2c_0 A_0), \hs{10} A_0 \equiv (2\pi)^2 R_0 R'
\eeq
where $f(g^2cA)$ is an unknown funtion of dimensionless parameter $g^2cA$. From this expression, we find that the partition function is independent of $\mu$. Therefore the limit $\mu \rightarrow \infty$ gives the exact partition function. In the discussion below, we assume that the partition function depends only on the area of torus $A$. We will check that the partition function Eq.\,(\ref{eq:partition function}) indeed gives the exact result for one vortex with $\NC=\NF=N$ and $\NF>\NC=1$.

We now perform the integration in (\ref{eq:partition function}) explicitly in some cases, all of which are new results.

\paragraph{1) Two ($\boldsymbol{k=2}$) local non-Abelian vortices with $\boldsymbol{\NC=\NF=2}$\\} 
Let us consider the domain of integration for $\NC=\NF=2$ and $k=2$ for example.
The domain can be easily seen from Fig.~\ref{fig:nc2nf2k2} and Fig.~\ref{fig:nc2nf2k2r}-(a).
In this case there are 
two rods with one particle inside each of them.
These rods can overlap with each other 
contrary 
to the case of $\NC=\NF=1$. 
However the edges of the rods cannot overlap with the particles inside the other rods. 
Therefore the domain of integration is given by
\beq
X'=\left\{(y_1,y_2,y_3,y_4) \left| \begin{array}{ccc} 0 < y_1 < 2 \pi R',& y_1 < y_2 < y_1 + d,& y_2 < y_3, \hs{5} y_1 + d < y_4,\\ y_3 < y_4 < y_3 + d,& y_3 + d < y_2 +2\pi R', & y_4 < y_1 + 2 \pi R', \end{array} \right. \right\}.
\label{eq:domain}
\eeq
Performing the integral in Eq.~(\ref{eq:partition function}) 
over the domain (\ref{eq:domain}),
we immediately obtain the partition function of the vortices 
(the volume of the domain wall (rods with particles) configuration space)
\beq
Z^{\NC=2,\NF=2}_{k=2,T^2} 
=  \left\{ \begin{array}{llc}
 \dfrac{1}{2} ( cT )^4  \left(\dfrac{4\pi}{g^2c}\right)^2
 A \left( A - \dfrac{2}{3}\dfrac{8\pi}{g^2c} \right)  
  &\qquad \mbox{ for } & \dfrac{8 \pi}{g^2c} \le A \\
 \dfrac{1}{6} ( cT )^4  \left(A-\dfrac{4\pi}{g^2c}\right)^2
 A \left( \dfrac{16\pi}{g^2c}  - A\right)  
  &\qquad \mbox{ for } &  \dfrac{4 \pi}{g^2c} \le A \le \dfrac{8 \pi}{g^2c} 
 \end{array} \right.. 
\label{eq:Z_k=Nc=Nf=2}
\eeq
{
Note that there exists a lower bound for the area $A \ge 4\pi/(g^2c)$
which corresponds to the Bradlow limit (\ref{eq:Bradlow}).   
}
This is a new result which was not derived previously.  
Let us explain the physical meaning of each factor 
restricting us to the first case.
Remembering that each vortex carries 
an internal orientation of $\mathbb{C}P^1$, 
we can understand that
the factor $(4\pi/(g^2c))^2$ corresponds to
the internal orientations (see Eq.~(\ref{eq:fdsa}) with $N=2$ {below}) 
while the factor $A$ does to the center of mass. 
Then the remaining factor 
$A - \frac{2}{3}\frac{8\pi}{g^2c}$ can be thought of as
effective area of relative motion moduli of two non-Abelian vortices. 
Comparing this result to
that of two Abelian vortices in Eq.~(\ref{eq:Z_k_ANO}) with $k=2$, 
the extra factor $2/3$ appears here, 
which implies that 
the effective area of non-Abelian vortices is smaller than
that of Abelian vortices. 
Interestingly, the factor 2/3 in the exclusion area is 
different from the factor 1/2 in the Bradlow area $B_2=\frac{1}{2}\frac{4\pi}{g^2c}$, contrary to the Abelian case Eq.~(\ref
{eq:eos}) in which the both factors coincide.
We will see this in more detail around (\ref{eq:pN}), below. 



\begin{figure}[h]
\begin{center}
\begin{tabular}{cc}
\includegraphics[width=65mm]{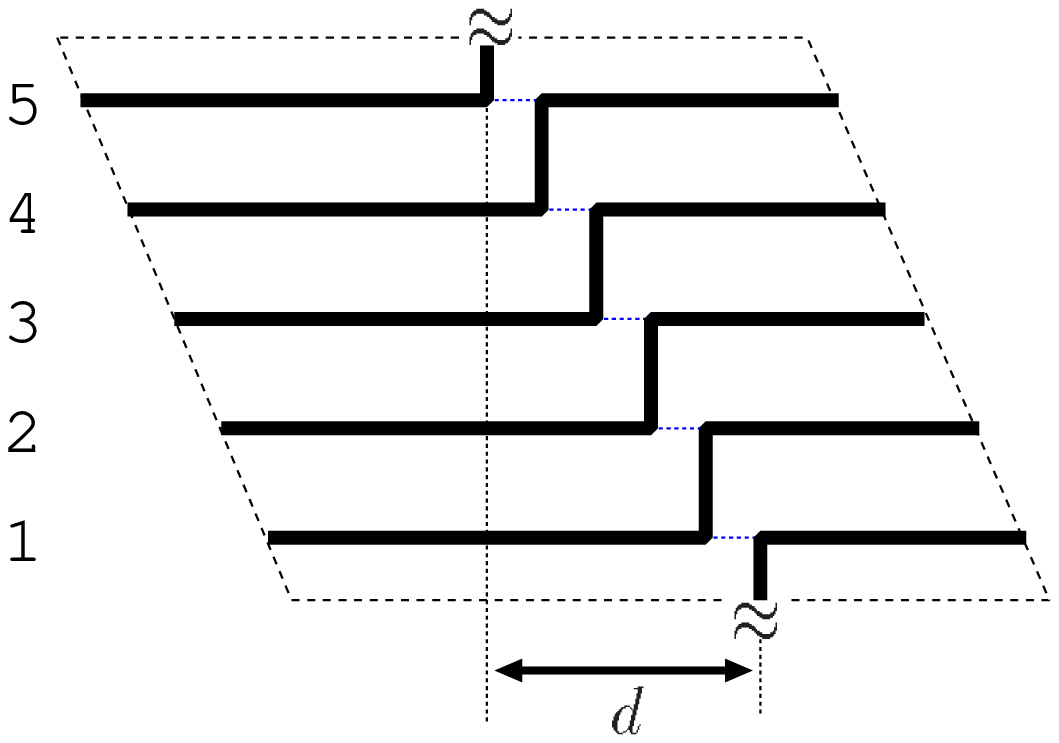}
 \hs{10} &
\includegraphics[width=60mm]{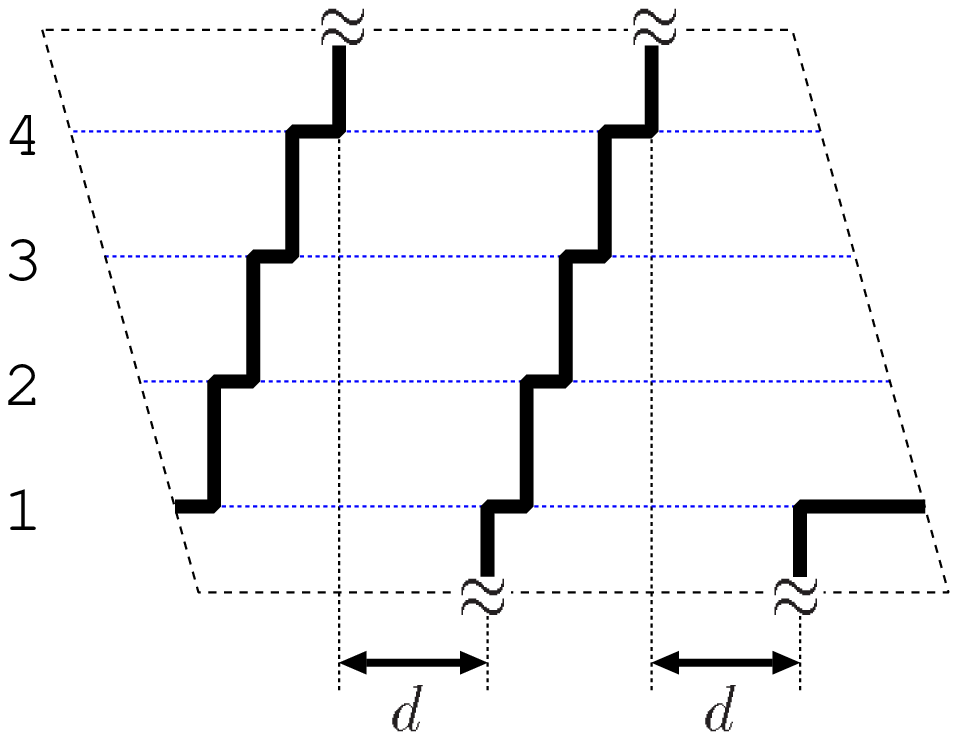} \\
(a) $\NC=\NF=5,~k=1$ & (b) $\NC=1,~\NF=4,~k=2$ \\ \\
\multicolumn{2}{c}{({\rm i}) 
Configurations of eigenvalues of $\hat \Sigma(x^1)$ on slant torus.
}
\end{tabular}
\begin{tabular}{cc}
\\
\includegraphics[width=60mm]{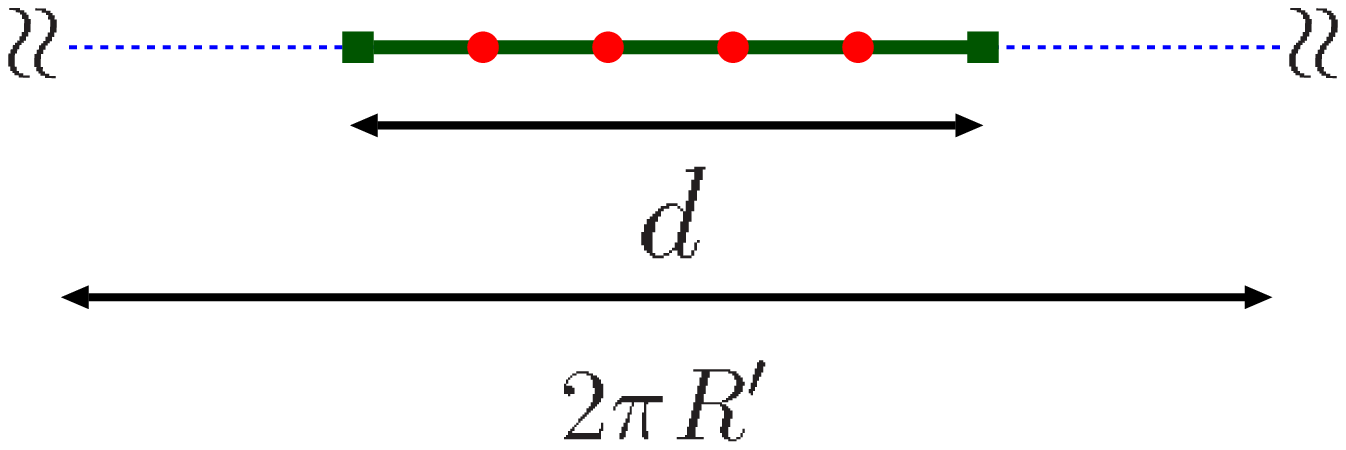}
 \hs{10} & 
\includegraphics[width=60mm]{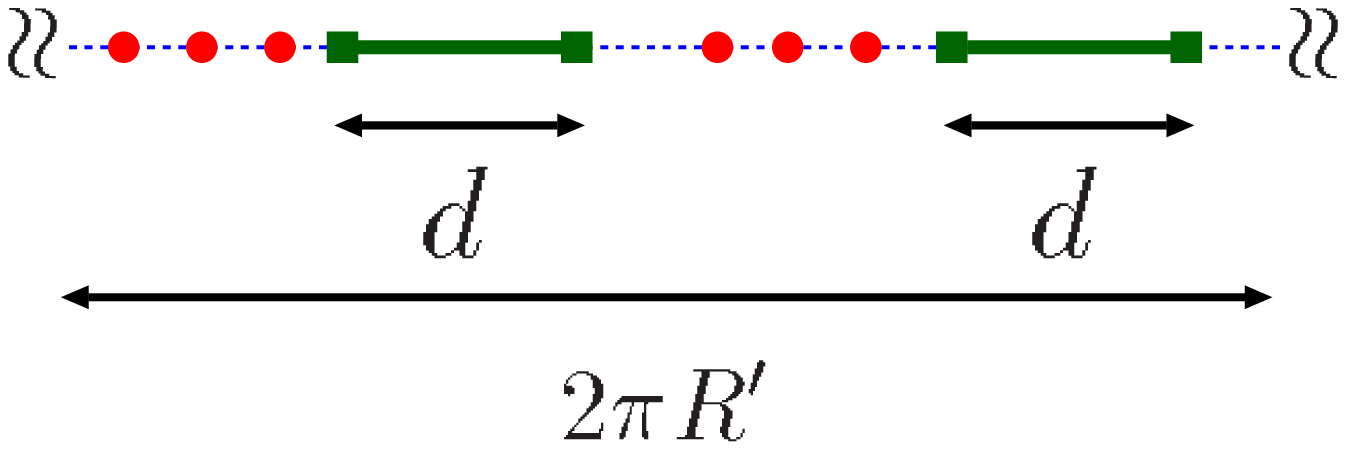} \\
(a) $\NC=\NF=5,~k=1$  & (b) $\NC=1,~\NF=4,~k=2$ \\ \\
\multicolumn{2}{c}{({\rm ii}) Configurations of rods and particles.}
\end{tabular}
\caption{\small 
Examples of stant configurations on eigenvalues of $\hat \Sigma(x^1)$ and corresponding rods and paricles. ({\rm i})-(a) There are 4 small kinks trapped in the region with length $d$. ({\rm i})-(b) There are 3 small kinks trapped in each interval between two regions with length $d$. 
({\rm ii})-(a) There are 4 particles trapped inside a rod. 
({\rm ii})-(b) There are 3 particles trapped in each interval between 2 hard rods.
}
\label{fig:fig6}
\end{center}
\end{figure}

\paragraph{2) $\boldsymbol{k}$ Local non-Abelian vortices with  $\boldsymbol{\NC=\NF=N}$\\ }
Let us next consider the single ($k=1$) non-Abelian vortex. 
In this case the moduli space is 
$T^2 \times \mathbb CP^{N-1}$. 
The corresponding configuration 
has a rod and $N-1$ particles trapped inside a rod, see
Fig.~\ref{fig:fig6}-(ii)-(a).

The partition function (\ref{eq:partition function}) is defined only in the region
$A\ge 4\pi/(g^2c)$ since 
the number of eigenvalues of $\hat \Sigma(x_1)$ is $N_{\rm C}$ 
and the period $2\pi R'$ should be larger than the length $d$ of a rod:
namely this bound is the second inequality in Eq.~(\ref{eq:Bradlow2}),
\beq
Z^{\NC=\NF=N}_{k=1,T^2} &=& 
\left( \frac{T}{2\pi} \right)^{N} (2\pi c)^{N} (2\pi R)^{N} 2 \pi R' \frac{1}{(N-1)!} d^{N-1}
\nonumber\\
&=& \frac{1}{(N-1) !} \left( cT \right)^{N} A \left(\frac{4\pi}{g^2c} \right)^{N-1}.
\label{eq:fdsa}
\eeq

In this simple case, we can confirm that our result agrees 
with an explicit integration over the exact metric
{on the non-Abelian vortex moduli space.}
Although the solution is not known, 
the metric with the K\"ahler class can be 
calculated \cite{Shifman:2004dr,Eto:2004rz}, to give
\beq
ds^2 = 2 \pi c \, dz d \bar z 
+ \frac{8\pi}{g^2} \frac{\delta_{i\bar j} 
(1+|b^k|^2) - \bar b ^i b^j}
{\left(1 + |b^k|^2 \right)^2} d b^i d \bar b^j,
 \label{eq:metric1}
\eeq
where $b^i$ are the inhomogeneous coordinates of 
{orientational moduli} $\mathbb CP^{N-1}$. 
Using this metric, we can confirm that Eq.~(\ref{eq:fdsa}) 
agrees with the correct partition function.


For general number $k$ of the non-Abelian vortices 
with $\NC=\NF=N$, we can calculate 
the  partition function (\ref{eq:partition function}) up to 
the next leading term in the expansion in terms of $1/A$
\beq
Z^{\NC=\NF=N}_{k,T^2} 
=  \left( cT \right)^{kN} \frac{1}{k!} 
\left[\frac{A}{(N-1)!}\left(\frac{4\pi}{g^2c}\right)^{N-1}\right]^{k}
\left[ 1 - D_N (k-1) \frac{k}{A}  +
\mathcal O\left(\left(\frac{4\pi}{g^2c A}\right)^{2}\right) \right].  
\label{eq:general-kN}
\eeq
We can show that the coefficient $D_N$ in Eq.~(\ref{eq:general-kN}) takes the form
(the definition of $D_N$ is given in Appendix)
\beq
 \frac{D_N}{4\pi/g^2c} = \frac{(2N-2)!!}{(2N-1)!!} = 1, \; {2\over 3}, \; {8\over 15}, \;
    {16 \over 35}, \; {128 \over 315}, \cdots \; 
\quad (N=1,2,3,4,5,\cdots) .  
 \label{eq:pN}
\eeq
The first (leading) term in the partition function 
(\ref{eq:general-kN}) represents the situation that 
all vortices are separated, 
while the second (next leading) term 
implies that a 
pair 
of adjacent vortices is overlaped. 
From the partition function (\ref{eq:general-kN}), 
we can obtain the equation of state for {dilute} vortex gas 
in the thermodynamic limit $A \rightarrow \infty, k\rightarrow \infty$ with
{
fixing $k/A$ to enough small value} as
\beq
 P A \left(1 - D_N \frac{k}{A}
 + {\cal O}\left(\frac{k^2}{A^2}\right) \right)=kT .
\eeq
From this we see that 
the second virial coefficient $D_N$ represents  
the effective area of the non-Abelian vortices in dilute gas. 
We find that the third term $\mathcal O(k^2/A^2)$ does not vanish. 
This implies that the equation of state deviates from van der Waals one, contrary to the Abelian case Eq.~(\ref{eq:Z_k_ANO}).

We find that $D_N < D_1 = \frac{4\pi}{g^2c}$ and $D_N$ begaves as
\beq
D_N \sim \frac{1}{2} \sqrt{\frac{\pi}{N}} \frac{4\pi}{g^2c}
\eeq
for large $N$. We see that the non-Abelian vortices can be closer to each other than the Abelian vortices even though the area of individual vortex is $\frac{4\pi}{g^2c}$ for both non-Abelian and Abelian vortices. Therefore we conclude that \textit{non-Abelian vortices are ``softer" than Abelian vortices.}

Clearly the inequality $B_N < D_N$ holds 
with $B_N=\frac{1}{N}\frac{4\pi}{g^2c}$ 
being the Bradlow area in (\ref{eq:Bradlow}). 
This inequality implies that 
the effective area $D_N$ of a vortex in dilute vortex gas 
is larger than the Bradlow area $B_N$, which is the effective area in highly pressured gas.
An intuitive understanding of this inequality is as follows.
It is known that the internal orientations of two non-Abelian vortices
are almost always aligned when the two vortices are approaching each other
\cite{Eto:2006db} (as long as their speed is sufficiently slow). 
In other words, nearby non-Abelian local vortices
behave as if they are Abelian local vortices.


\paragraph{3) $\boldsymbol{k}$ semi-local vortices with $\boldsymbol{\NC=1}$ and general $\boldsymbol{\NF}$\\}
Finally we show that our method can be 
extended to the case of semi-local vortices with 
$N_{\rm F} > N_{\rm C}$. 
Here we concentrate on the $N_{\rm C}=1$ case 
which indicates essential features of semi-local vortices.
The corresponding configuration has
$\NF-1$ particles trapped in each interval between $k$ hard rods,
see Fig.~\ref{fig:fig6}-(b).
In this case, Eq.~(\ref{eq:partition function}) reduces to
\beq
Z^{\NC=1,\NF}_{k,T^2} 
 &=& \left( \frac{T}{2\pi} \right)^{k\NF} 
 (2\pi c)^{k\NF} (2\pi R)^{k\NF} \frac{1}{k} \frac{1}{(k\NF-1) ! } 
 2 \pi R' \left( 2\pi R' - d k \right)^{k\NF-1} \nonumber \\ 
&=& \left( cT \right)^{k\NF} \frac{1}{k} \frac{1}{(k\NF-1) ! } 
A \left( A-\frac{4\pi k}{g^2c} \right)^{k\NF-1},
\label{eq:asdf}
\eeq
where $1/k$ factor is needed since $k$ vortices cannot be distinguished.
Substituting $\NF=1$ into this result, the partition
function of $k$ ANO vortices in Eq.~(\ref{eq:Z_k_ANO}) is correctly recovered. 
From this partition function, we can obtain the equation of state for the vortex gas in the thermodynamic limit $A\rightarrow \infty,k\rightarrow \infty$ with $k/A$ finite as
\beq
P\left(A - \frac{4\pi k}{g^2c} \right) = k \NF T,
\label{eq:eosNf}
\eeq
where $P$ is the pressure. 
The factor $\NF$ in the right hand side of 
Eq.~(\ref{eq:eosNf}) appears due to the fact that the 
vortices have additional internal degrees of freedom in 
the case of $\NF>1$.

In this case, no field theoretical result has been known yet. 
Especially we can show that 
the moduli space of $k=1$ semilocal vortex (see Fig.~\ref{fig:fig6}-(b))
is $T^2 \times {\mathbb C}P^{\NF-1}$ 
by extending the moduli matrix formalism \cite{Eto:2006pg} 
to $T^2$.
Second, as was done in \cite{Eto:2004rz}, we can compute  
the moduli space metric by using the integration formula 
of the K\"ahler potential in \cite{Eto:2006uw}, 
to give 
\beq
ds^2 = 2 \pi c \, dz d \bar z 
+ 2 c \left( A - \frac{4\pi}{g^2c} \right) 
\frac{\delta_{i\bar j} (1+|b^k|^2) - \bar b ^i b^j}
{\left(1 + |b^k|^2 \right)^2} d b^i d \bar b^j.  \label{eq:metric2}
\eeq
Here $b^i$ in the second term are the inhomogeneous coordinates of 
$\mathbb CP^{\NF-1}$ and represent moduli not localized around vortices.
These moduli become non-normalizable on the plane $\mathbb C$ ($A \rightarrow \infty$). However they contribute to the K\"ahler potential since we are considering
the vortices on  the compact manifold $T^2$ with the finite area $A$.
From this metric we can confirm at least for $k=1$ case 
that Eq.~(\ref{eq:asdf}) 
gives the correct partition function.


\section{Comments and Discussions}

We here comment on an important fact that there is a duality on the partition
function (\ref{eq:partition function}), 
without 
performing 
the difficult integration. 
That is, the partition function  
for $k\le \NC$ is invariant under exchange
\beq
&&\left(\NC,\, g^2\right) \leftrightarrow 
\left(\widetilde N_{\rm C} ,\,\tilde g^2 \right),\\\nn
{\rm with~}&& \widetilde N_{\rm C} \equiv k + \NF -\NC,\quad 
{ 4\pi \over \tilde g^2 c} \equiv A - {4\pi\over g^2c}.
\eeq
One can confirm that this duality exists between 
the partition function (\ref{eq:fdsa}) and 
the one of the $k=1$ case of (\ref{eq:asdf}). 
Also, the example of $\NC=\NF=2, k=2$ is self-dual. 
In fact, in the partition function (\ref{eq:Z_k=Nc=Nf=2}), 
the latter case ($A/2 \le 4\pi/(g^2c)$)  
can be obtained by exchanging 
$4\pi/(g^2c)$ by $A - 4\pi/(g^2c)$ in the former case 
 ($A/2 \geq 4\pi/(g^2c)$).
These observations are very analogous to the duality 
relation between the non-commutative 
instantons and  monopoles 
\cite{Hashimoto:1999zw,Gross:2000wc}. 
From string theoretical point of view, there appears a 
slant torus due to the effect of the magnetic gauge field 
or equivalently the $B$-field. The $B$-field is a source 
of the non-commutativity and associated with the FI 
parameters in the effective 
theory on the instantons.
In our case, the gauge coupling plays a role of the 
non-commutative parameter and should relate to the 
modulus of the slant torus. 
Better explanation on the above duality may be found 
in terms of the duality between the non-commutative vortex 
and domain wall in the string theoretical picture. 

\bigskip
Finally we would like to discuss vortices on a sphere $S^2$. 
In this case  
we cannot use the 
above T-dual argument 
naively 
since the sphere consists of non-trivial $U(1)$ fibration. 
However we here assume that the vortices on the sphere is equivalent to the
vortices on the cylinder 
with finite length 
in the computation of the volume of the moduli space; 
The cylinder is topologically isomorphic to the sphere 
with two puncture points.
When at least one vortex sits on a puncture point 
(the notrh or south pole) of the sphere, 
the dimension of the corresponding subspace is less than 
the dimension of the full moduli space. 
Therefore this subspace does not contribute to 
the volume of the moduli space. 
In contrast 
to 
the torus case, 
we do not need to 
identify the both sides in the $x^1$-direction.
Thus it is now sufficient to consider the gas of the hard rods in a finite segment with length $2\pi R'$.
By using the system of the hard rods, 
we obtain the partition function of vortices in the Abelian-Higgs model 
on $S^2$ as 
\beq
Z^{\NC=\NF=1}_{k,S^2} = \frac{1}{k!} \left(c T \right)^k \left( 2\pi R' - kd \right)^k \left( 2\pi R\right)^k = \frac{1}{k!} \left(c T \right)^k \left( A - \frac{4 \pi k}{g^2c} \right)^k.
 \label{eq:S^2}
\eeq
This completely agrees with the results in 
\cite{Manton:1993tt,Manton:1998kq,Manton:2004tk}. 
This case also should be extendable to the non-Abelian and/or semi-local 
vortices.

\medskip
In conclusion,
we have proposed a novel and simple 
method to compute the partition function of 
vortices at finte temperature. 
Our result agrees with previously 
known cases (\ref{eq:Z_k_ANO}) and (\ref{eq:S^2}) 
of the local Abelian (ANO) vortices in the Abelian-Higgs model on 
$T^2$  and $S^2$, respectively. 
Our method provides new results in more general cases 
of non-Abelian local vortices, 
(\ref{eq:Z_k=Nc=Nf=2}) and (\ref{eq:general-kN}),
and Abelian semi-local vortices, (\ref{eq:asdf}).
In the two cases of $k=1$, $\NC=\NF=N$ and 
$k=1$, $\NC=1$ with general $\NF$, 
we have confirmed that our results agree 
with explicit integrations over  
the exact moduli metrics
(\ref{eq:metric1}) and (\ref{eq:metric2}), respectively.
We have found that non-Abelian vortices 
are reduced under T-duality to soft rods 
with particles inside them while 
the Abelian vortices are to hard rods.

Our results 
will be applied to 
the thermal  vortex gas in the early Universe or 
in superconductor. 
Extension to non-(or near-)critical coupling 
will be important to discuss more realistic application.
In particular the phase transition 
of vortex gas in the Abelian-Higgs model 
was discussed previously \cite{Shah:1994sx,Kajantie:1999ih}. 
Presence or 
absence 
of phase transitions in 
the non-Abelian and/or semi-local vortices  
is very interesting to explore.

\section*{Acknowledgements}

This work is supported in part by Grant-in-Aid for 
Scientific Research from the Ministry of Education, 
Culture, Sports, Science and Technology, Japan 
No.17540237and No.18204024 (N.~S.). 
K.~Ohta is supported in part by the 21st 
Century COE Program at Tohoku University
``Exploring New Science by Bridging Particle-Matter Hierarchy.''
The work of K.~Ohashi. and M.~E. is
supported by Japan Society for the Promotion 
of Science under the Post-doctoral Research 
Program. 
T.~F. gratefully acknowledges 
support from a 21st Century COE Program at 
Tokyo Tech ``Nanometer-Scale Quantum Physics" by the 
Ministry of Education, Culture, Sports, Science 
and Technology, 
and support from the Iwanami Fujukai Foundation. 
The authors thank the Yukawa Institute for Theoretical  
Physics at Kyoto University. Discussions during the YITP workshop  
YITP-W-06-11 on ``String Theory and Quantum Field Theory'' were  
useful to complete this work. K.~Ohta would like to thank S.~Matsuura for useful discussions and comments.


\appendix
\section{Virial Expansion}
In this appendix we give the defenition of $D_N$ which appears
as a coefficient of the next leading term in the partition function 
(\ref{eq:general-kN}).
We consider $k$ non-Abelian local vortices with $\NC = \NF = N$ on a torus $T^2$.
As we have mentioned, the system can be well described
from the picture of $k$ soft rods with length $d$, 
each of them piercing $N-1$ particles therein, on $S^1$ in circumference 
$L=2\pi R'$,
as shown in Fig.~\ref{fig:k-vortex}. 
We can carry out an integration with respect to only $k$ parameters
corresponding to positions of the rods and we find that     
the volume of the configuration space can be rewritten as
\begin{eqnarray}
 V_{N,k}\equiv \frac{1}k\int d^{kN}y=\frac{\left(Ld^{N-1}\right)^k}{k!}\int_{\Sigma_{N,k}} d^{k(N-1)}z
\left[\max\left(1-\frac{d}L \sum_{i=1}^kM_{N,i}\,\,,\,0 \right)\right]^{k-1},
\label{VNk}
\end{eqnarray}
where the integration parameters $z=z_{r,i}, (r=1,\cdots,N-1, i=1,\cdots,k)$ 
are dimensionless and
correspond to $(N-1)$ relative positions of the particles
inside each rod, and their integration region is defined by
$\Sigma_{N,k}=
\left\{\{z_{r,i}\}\big|0\le z_{1,i}\le z_{2,i}\le\cdots\le z_{N-1,i}\le 1,
1\le i\le k \right\}$.
Here we have defined a dimensionless function, $M_{N,i}$, for each rod
($1\le i\le k$),
\begin{eqnarray}
M_{N,i} &=& \max(z_{1,i},z_{2,i}-z_{1,i+1},\cdots,z_{N-1,i}-z_{N-2,i+1},1-z_{N-1,i+1}),
\label{eq:MNi}
\end{eqnarray}
where we have defined $z_{r,k+1} = z_{r,1}$. All of them give the same contributions to the integration. 
Note that dimensionless parameter $d/L$ appears in the integrand.
We find $kd \geq d \sum_i M_{N,i}$ in this case of local vortices. Thus small $kd/L$ guarantees that $L \geq d \sum_i M_{N,i}$ in any point of the integration region and that the integrand can be expanded with respect to $d/L$, since the max-function in Eq.~(\ref{eq:MNi}) can be ignored\footnote{
In the case of semilocal vortices, $N_{\rm F}>N_{\rm C}>1$,
we cannot take the similar operation since 
we cannot naively remove the max-function from the integral representation (\ref{VNk})
due to their size moduli integral.}.
The $D_N$ appears as the coefficient of the next leading term 
in this expansion of $V_{N,k}$,
namely in the virial expansion of soft rods in one dimension:
\begin{eqnarray}
 V_{N,k}=\frac{1}{k!}\left(\frac{Ld^{N-1}}{(N-1)!}\right)^k
\left(1-\hat{D}_N\times  k(k-1)\frac{d}L+ {\cal O}\left(\frac{d^2}{L^2}\right)\right),
\end{eqnarray}
where we have defined $\hat{D}_N$ as an average value of $M_{N,1}$:
\begin{eqnarray}
\hat{D}_N 
&=& 
\frac{1}{N} + \left((N-1)!\right)^2
\int_{\substack{
\\\\
0 \le z_1 \le \cdots \le z_{N-1} \le 1 \\
0 \le z'_1 \le \cdots \le z'_{N-1} \le 1 
}} 
\!\!\!\!\!\!\!\!\!\!\!\!\!\!\!\!\!\!\!\!\!\!\!\!\!\!\!\! 
dz^{N-1}
dz^{'N-1}\ 
\max\left(z_1-z_1',\cdots,z_{N-1}-z'_{N-1},0\right),
\end{eqnarray}
where $\hat{D}_N$ relates with $D_N$ by $D_N=\hat{D}_N \frac{4\pi}{g^2c}=\hat{D}_N2\pi R d$.
We can easily carry out this integration  
in each region where a definite order of the integration
parameters such as $z_1<z_1'<z_2'<z_2<\cdots$.
It is convenient to divide the integration regions into several ones whose integral become identical.
Therefore calculation for $D_N$ reduces to counting number of elements 
of the sets and we have succeeded this counting and obtained the result
as (\ref{eq:pN}).



\end{document}